\documentclass[%
aps,
prb,
 amsmath,
 amssymb, 
reprint,%
numerical,
]{revtex4-1}
\usepackage{epstopdf} 
\usepackage{color}
\usepackage[english]{babel}
\usepackage{amsmath}
\usepackage{amsfonts}
\usepackage{graphicx}
\usepackage{array}

\begin{document}

\title{Two-dimensional currents at semiconductor surfaces as resonances within the classical current equation}
\author{Jakub Lis}
\email{j.lis@uj.edu.pl}
\affiliation{Center for Nanometer-Scale Science and Advanced Materials (NANOSAM), Faculty of Physics, Astronomy and Applied Computer Science, Jagiellonian University, ul. St.  Lojasiewicza 11, 30-348 Krakow, Poland}

\begin{abstract}   
Logarithmic voltage profile characteristic to two-dimensional current flows  has been reported for several semiconductor surfaces.  We analyze this phenomenon within a simple model of accumulation and inversion layers.  Prompted by numerical analysis of several conductivity profiles we point out how the logarithmic voltage drop may appear in the case of three-dimensional samples. It may or may not be observed both for the accumulation and the inversion layers.   Further, the numerical data suggest   that  the surface conductivity at large distances is governed by very few parameters.
\end{abstract}

\maketitle
Electric current measurements at surfaces at sub-micron scale have become increasingly popular\cite{popular}.
There have been several reports on observation of two-dimensional current flow at semiconductor surfaces\cite{APL,JAP,2d1,2d2,2d3}. It has been suggested that this is due to formation of an inversion layer near the semiconductor surface. Alternatively, in  cases when two-dimensional currents are observed at lower temperatures only,  decoupling from bulk conductivity  was argued, e.g. Ref.~\onlinecite{2d3}. There are no conclusive theoretical results available allowing evaluation of these intuitive concepts.
In our recent paper~\cite{PRB} we have developed a theoretical framework to understand how subsurface conductivity variation impacts the voltage profile at the surface. We used the classical current equation assuming proportionality between the current density and the electrostatic field without  regard to quantum surface  phenomena. It was shown that the actual surface conductivity can be measured close to the current source only while at large distances the bulk conductivity prevails. Here, we show that at intermediate distances the voltage drop may be described by the logarithmic function which is a fingerprint of the two-dimensional current flow. As such, the dimensional reduction emerges as an effective description suitable on restricted distances. The presence of an inversion layer is not a sine-qua-non  condition for its appearance which may also be facilitated by an enhancement of the conductivity near the surface (accumulation layer). \\
The equation we consider here was examined with mathematical rigor in the context of the so-called Calderon problem~\cite{Calderon}. That theory puts the electrical impedance tomography~\cite{EIT} on firm ground. Despite some overlap between our work and those developments, these two problems are both physically and mathematically autonomous.\\

The macroscopic  equation governing the current flow with a source $f$ has the following form
\begin{equation}\label{start}
\nabla\left(\sigma\nabla\Phi\right)=f,
\end{equation}
where $\Phi$ is the electrostatic field,  $\sigma$  -- the conductivity and $\nabla$ -- the nabla differential operator. The current density $\mathbf{j}$ may be calculated from the relation
\begin{equation}
\mathbf{j}(\mathbf{x})=\sigma(\mathbf{x}) \nabla \Phi(\mathbf{x}).
\end{equation}
We aim at finding the surface voltage profile at the sample-vacuum boundary. So, we solve the equation in a semi-space, i.e with two unrestricted coordinates ($x$ and $y$) and the third one $z>0$; the surface is located at $z=0$. The boundary condition excludes any current flowing through the surface, hence
\begin{equation}\label{boundaryEq}
\sigma(0)\frac{\partial\Phi}{\partial z} =0.
\end{equation}
The conductivity is assumed to depend on the distance from the surface only $\sigma=\sigma(z)$, and to reach a constant value $\sigma(\infty)$ in the bulk. The variation of $\sigma$ may be due to the Fermi level pinning phenomenon known in the semiconductors or a specific surface treatment.  The model abandons the tip induced change of the conductivity, for more detail see~\cite{PRB}. In that paper it was pointed out that the theory may result in a strange long-ranged behavior if $\tfrac{d}{dz}\sigma(0)\neq 0$. We clarify  this issue in Appendix~\ref{app} concluding that the condition can be treated on an equal footing with the condition $\tfrac{d}{dz}\sigma(0) = 0$.

To arrive at the general solution of eq.~(\ref{start}) we look for solutions of function $\xi$
\begin{equation}\label{transform}
\xi=\sqrt{\sigma} \Phi,
\end{equation}
which recasts eq.~(\ref{start}) into a Schr{\"o}dinger-like form
\begin{equation}
\underbrace{\left[-\Delta+\frac{\Delta \sqrt{\sigma}}{\sqrt{\sigma}}\right]}_{\hat{L}}\xi=\sigma^{-1/2}(x)f.
\end{equation}
The boundary condition~(\ref{boundaryEq}) reads
\begin{equation}\label{boundaryTransformed}
\left.\frac{d\xi}{dz}\right|_{z=0}=\left.\frac{1}{2\sigma}\frac{d\sigma(z)}{dz}\right|_{z=0}.
\end{equation}
The useful formula for the Green function $G$ is given in terms of solutions of the following one-dimensional equation 
\begin{equation}\label{1dim}
\left(-\frac{d^{2}}{dz^{2}}+V(z)\right)\psi(k;z)=k^{2}\psi(k;z),
\end{equation}
where the potential $V(z)$  is given by formula
\begin{equation}
V(z)=\sigma^{-1/2}\frac{d^{2}\sqrt{\sigma}}{dz^{2}}.
\end{equation}
The general solution of eq.~(\ref{start}) reads 
\begin{equation}
\Phi(\mathbf{x})=\int dx'dy'dz' G(\mathbf{x};\mathbf{x'})f(\mathbf{x'}),
\end{equation}
where 
\begin{equation}
\begin{split}
G(\mathbf{x};\mathbf{x'})=\frac{1}{2 \pi \sqrt{\sigma(z)\sigma(z')}} \\ \times  \int_{0}^{\infty} dk \  K_{0}(k\sqrt{(x-x')^{2}+(y-y')^{2}})\psi(k;z)\psi(k,z').
\end{split}
\end{equation}
The model considered below involves a point source located at  $(x'', y'' ,z''=0)$. Hence,   we formally write
\begin{displaymath}
f(\mathbf{x'})=I\delta(x'-x'')\delta(y'-y'')\delta(z')
\end{displaymath}
to arrive at the formula for the surface voltage profile $\phi(r)=\Phi(r,z=0)$ normalized by the current $I$ supplemented by the source
\begin{equation}\label{integ}
I^{-1}\phi(r)=\frac{1}{2\pi\sigma(0)}  \int_{0}^{\infty} dk \  K_{0}(kr)\psi^{2}(k;0),
\end{equation}
where $r$ stands for the two-dimensional radius $\sqrt{(x-x'')^{2}+(y-y'')^{2}}$.  \\

First, we comment on a natural conjecture that the two-dimensional currents are due to bound state solutions of eq.~(\ref{1dim}). To this end,  we write the expression for energy  released  by the current in the source-free region of space per unit time. It reads~\cite{Landau}
\begin{equation}\label{en1}
E[\Phi]=\frac{1}{2} \int  d\Omega \ \sigma \left(\nabla\Phi\right)^{2},
\end{equation}
where the integration is done over the whole region and $\Phi$ stands for the electrostatic potential. Note that eq.~(\ref{start}) in the source-free region is equivalent to finding extremum of the above energy functional. For any given function $\Phi$, the functional results in a non-negative number in agreement with its physical interpretation. Upon substitution~(\ref{transform}) the energy  gains an  equivalent form
\begin{equation}\label{en2}
E[\xi]=\frac{1}{2} \int  d\Omega \ \xi\hat{L}\xi
\end{equation}
 excluding any negative eigenvalue as well as the zero eigenvalue with a square-integrable eigenfunction. The equivalence between functionals~(\ref{en1}) and~(\ref{en2}) can be shown if boundary condition~(\ref{boundaryTransformed}) is taken into account.   As the potential $V(z)\to 0$ for $z\to\infty$, we conclude that no bound states are allowed and only continuous (scattering) spectrum is present. This demonstrates that the conclusions of Ref.~\cite{PRB} are valid in any case, i.e., the current will have the three-dimensional character  at sufficiently small and large distances from the  source. So, the confinement close to the surface may appear  on limited distances only. To get the idea how the function $\psi^{2}(k;0)$ looks like we numerically analyze several conductivity profiles. Only then we can comment on the mechanism behind dimensional reduction.\\

For the presentation of the surface voltage profile we will stick to the scheme explored in recent experimental reports~\cite{APL,JAP}.
It follows from the experimental geometry where two current supplying electrodes are put at the surface at $(0,0)$ and $(D,0)$ and two additional tips measure the voltage drop between the points $\tfrac{D}{2}(1+x,0)$ and $\tfrac{D}{2}(1-x,0)$ for some $0<x<1$. The resistance $R$ (voltage drop divided by the current flowing through the system) is shown as a function of $x$ for several $D$. It is known~\cite{APL} that the measured resistance $R$  depends on $x$ only in the case of two-dimensional currents. Both $x$ and $D$ are necessary  to determine the resistance in the three-dimensional case where the quantity $R\cdot D$  approximately  is independent of  $D$\cite{APL,PRB}. In both cases the formulas for the resistance are analytic and yield
\begin{equation}\label{2d}
R_{2}=\frac{1}{\pi\sigma_{2}}\ln\frac{1+x}{1-x}
\end{equation}
for two-dimensional currents, where $\sigma_{2}$ is a two-dimensional conductivity parameter and 
\begin{equation}\label{3d}
R_{3}=\frac{1}{ D\pi\sigma}\frac{x}{1-x^{2}}
\end{equation}
for three dimensional-currents.\\

\begin{figure}
\includegraphics[scale=0.35]{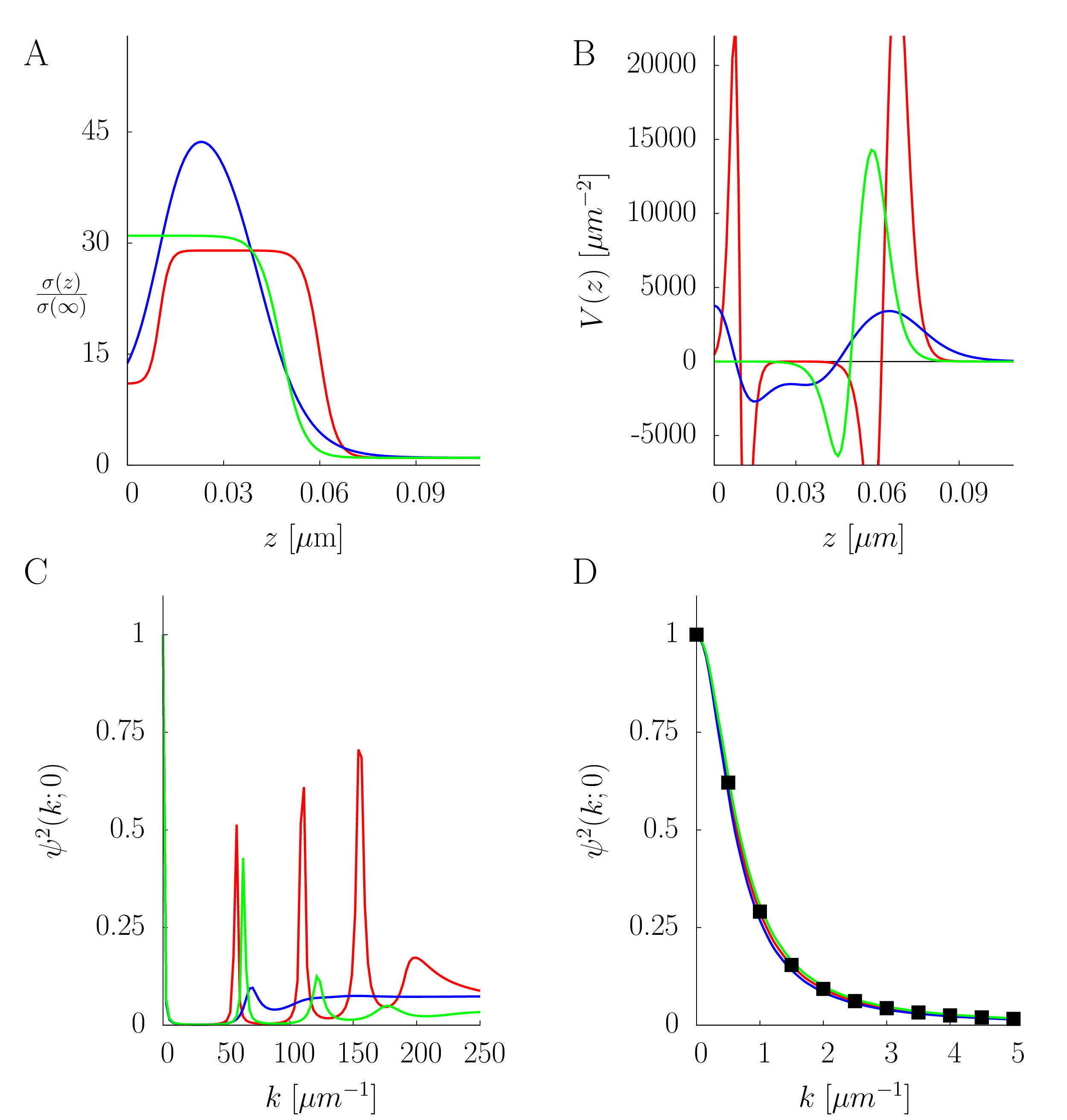}
\caption{Panel A shows three conductivity profiles numerically considered  $ 14(1-\tanh \tfrac{x-0.06}{0.005})-9(1-\tanh \tfrac{x-0.01}{0.003})+1$ -- red, $25(1-\tanh \tfrac{x-0.0405}{0.015})-21(1-\tanh \tfrac{x-0.01}{0.01})$ -- blue, and $15(1-\tanh \tfrac{x-0.048}{0.007})+1$ -- green. The resulting potentials $V(z)$ are shown in panel B; following functions $\psi^{2}(k;0)$  in  panel C. Image D is a close view of $\psi^{2}(k;0)$ for small $k$ with several points corresponding to the Lorentzian as defined in eq.~(\ref{lorentz}) with $\Gamma^{2}=0.41 \ \mu m^{-2}$, $k_{0}=0$ and $A=1$ shown for illustration.  Functions $\psi^{2}(k;0)$ are rescaled  to satisfy $\psi^{2}(0;0)=1$ (panels C and D).\label{potent}}
\end{figure}
\begin{figure*}[!]
\includegraphics[scale=0.45]{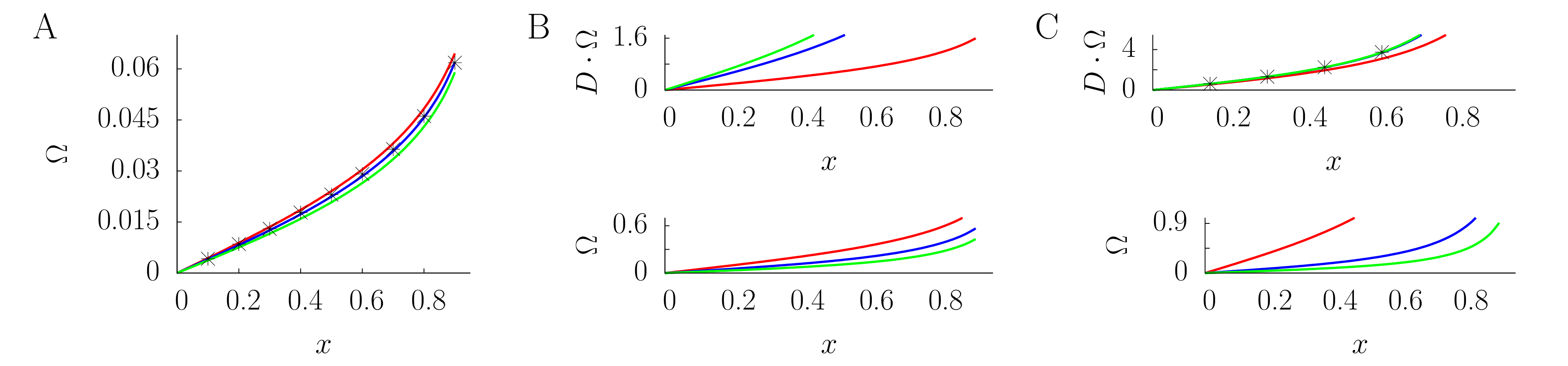}
\caption{Normalized resistance $\Omega= \sigma(\infty) \phi/I$ for the experimental geometry for various values of parameter $\Gamma^{2}$: $\Gamma^{2}=0.0005 \ \mu m^{-2}$ in A, $\Gamma^{2}=0.1 \ \mu m^{-2}$ in  B, and $\Gamma^{2}=20 \ \mu m^{-2}$ in C. Colours correspond to different values of $D$: $2\ \mu m$ -- green, $10\ \mu m$ -- blue, and $20\ \mu m$ -- red. Two-dimensional character resulting in independence of $D$ is shown in panel A. Nearly three-dimensional case is shown in panel C as the quantity $DR\sigma(\infty)=D\Omega$ appears $D$-independent. An examplary transition behavior is shown in panel B, where neither graphs of $\Omega(x)$ nor $D\Omega(x)$ coincide. The points (stars) in panels A and C demonstrate the validity of eq.~(\ref{2d}) and eq.~(\ref{3d}), respecively. \label{resist}}
\end{figure*}

We analyze three  close  but clearly distinct conduction profiles, see Fig.~\ref{potent}A. They are motivated by investigations of TiO$_{2}$  surface conducting layer reported in Ref.~\onlinecite{Rogala} and  measurements performed in our group (not published).  Cycles of sputtering and heating of the sample  result in creation of a thin (30-40~nm in width) layer with conductivity exceeding its bulk value. The profiles are shown in Fig.~\ref{potent}A, and the resulting potentials  $V(z)$ in Fig.~\ref{potent}B. The function $\psi^{2}(k;0)$ has several pronounced peaks as demonstrated in  Fig.~\ref{potent}C. 
It turns out that for small $k$ all functions $\psi^{2}(k;0)$ appear barely distinguishable around $k=0$ and may be well approximated by a Lorentzian. According to our numerical experience it is a generic behavior: any well-defined peak of $\psi^{2}(k;0)$  is satisfactorily reproduced by  Lorentz function
\begin{equation}\label{lorentz}
g(\Gamma,A,k_{0};k)=\frac{A\Gamma^{2}}{(k-k_{0})^{2}+\Gamma^{2}},
\end{equation}
where $A$ stands for the peak  amplitude, $\Gamma^{2}$ for its width, and $k_{0}$ for its position. Discrepancies are observed for tails and they will not bother us here.  \\
According to  formula~(\ref{integ}), the surface potential $\phi(r)$  emerges upon integration.
Function $K_{0}(x)$ explodes at the origin and quickly vanishes starting from $x\approx 3$. This makes all other peaks visible in Fig.~\ref{potent}C irrelevant at experimentally accessible distances, roughly above 100~nm. Then we can approximate the function $\psi^{2}(k;0)$ by Lorentzian with $k_{0}=0$, $\psi^{2}(k;0)=g(\Gamma,A,0;k)$.  The amplitude $A=\psi^{2}(0;0)$ is a multiplicative prefactor which cannot alter the functional form of the voltage profile.
Plugging relation~\cite{PRB} $\psi^{2}(0;0)=\tfrac{\sigma(0)}{\sigma(\infty)}$ into eq.~(\ref{integ}) we arrive at an effective formula
\begin{equation}
I^{-1}\phi(r)\sigma(\infty)=\frac{1}{2\pi  r}\int_{0}^{\infty}dx \ K_{0}(x)g\left(\Gamma,1,0;\frac{x}{r}\right),
\end{equation}
which reduces the complexity of $\psi^{2}(k;0)$ to just one parameter: the width of the Lorentzian.
 It is demonstrated in Fig.~\ref{resist}  that such a  simplified function $\psi^{2}(k;0)$ can describe both three-dimensional and two-dimensional currents. In Fig.~\ref{resist} there are  shown curves obtained for several values of $\Gamma^{2}$. For small ones the numerically simulated  curves nearly coincide (Fig.~\ref{resist}A) which is a fingerprint of the two-dimensional conductance.  If the peak is broad,  three-dimensional currents are manifest as in Fig.~\ref{resist}C. Fig.~\ref{resist}B shows results for an intermediate value of $\Gamma^{2}$.\\

\begin{figure}
\includegraphics[scale=0.3]{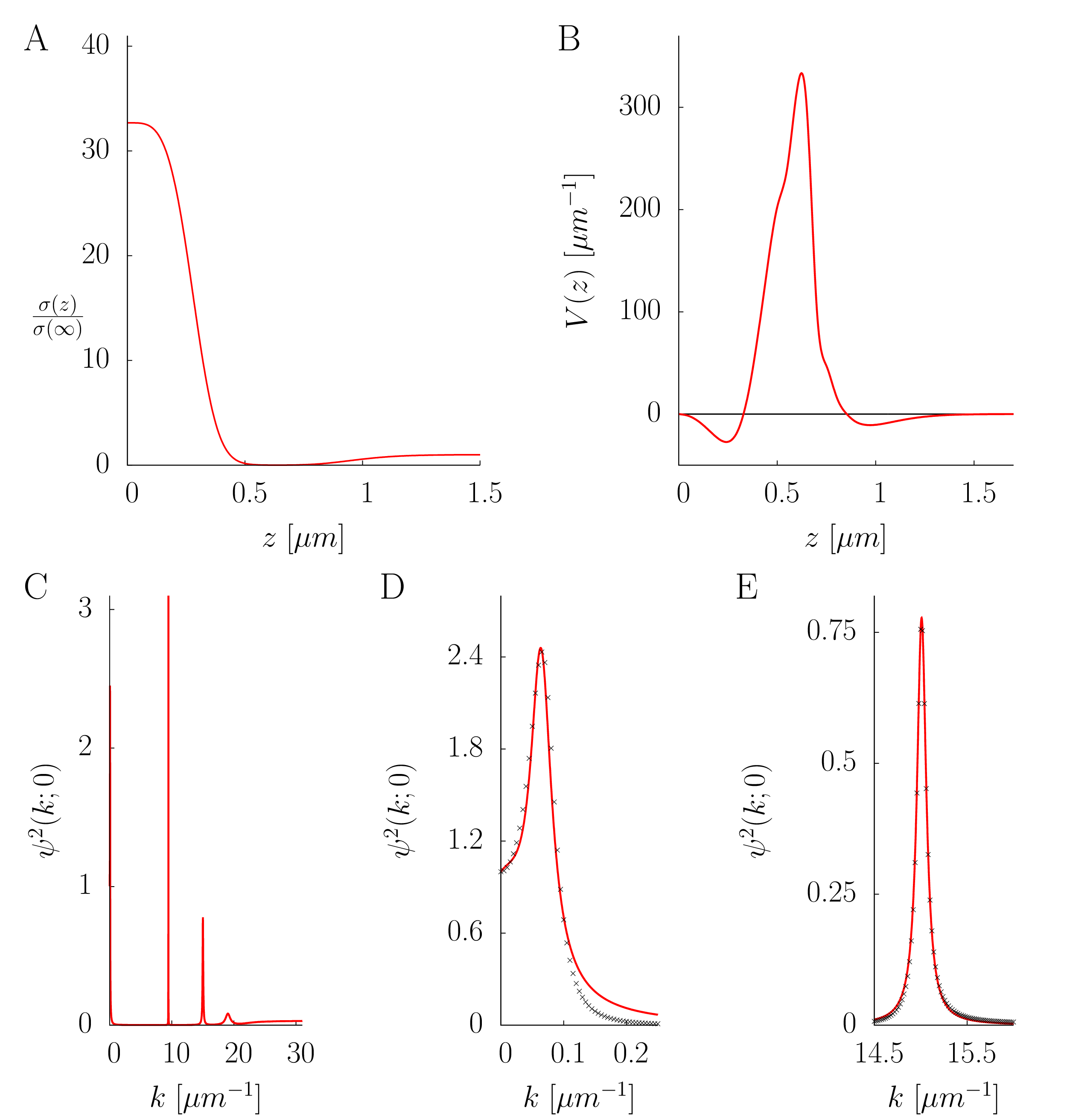}
\caption{Conductivity profile corresponding to an inversion layer with conductivity at the surface larger than the bulk one is shown in panel A. Resulting potential $V(z)$ is shown in panel B, while the following function $\psi^{2}(k;0)$  in  panel C. Images D and E  show close views of $\psi^{2}(k;0)$ for small $k$ (image D) and  around the third peak (image E). The points (stars) correspond to numerically calculated values, the solid line in D corresponds to a sum of two Lorentz functions and to a single Lorentzian  in E. \label{resist2}}
\end{figure}

In most reports, the dimensional reduction of the current is associated with the existence of an inversion layer~\cite{APL,JAP,2d1,2d2}. The  above analyzed $\sigma$ functions  do not model such systems which have a more complicated conductivity  profile, see Fig.~\ref{resist2}A. To arrive at such profile we use textbook  approximations~\cite{Monch}. We denote with $\eta(z)$ a monotonically decreasing function interpolating between the surface and bulk Fermi levels. The carrier density at some point $z$ is described as
\begin{equation}
n_{h/e}(z)=\tilde{n}_{h/e} e^{\pm \beta\eta(z)},
\end{equation}
where $\tilde{n}_{h/e}$ stands for the bulk density of holes/electrons. We assume that the positive sign in the exponent holds for holes and negative for electrons.
The conductivity may be written
\begin{equation}
\sigma(z)=|n_{e}|v_{e}+n_{h}v_{h},
\end{equation}
where  $v_{e}$ and $v_{h}$ stand for electron and hole mobility, respectively. If the mobility values are of the same order, than    the plotted function $\sigma(z)$ is generic with a single reservation: $\sigma(0)/\sigma(\infty)$ can be both greater and smaller than unity. As shown in Fig.~\ref{resist2}C  resulting $\psi^{2}(k;0)$ has a minimum at the origin and a peak for some $k_{0}$. Despite this difference from the above described shapes, for given scale $D$ it can produce all surface potential types shown in Fig.~\ref{resist}. \\

To unravel the mechanism behind appearance of the logarithmic potential profile we analyze the integral formula~(\ref{integ}). It is known that for $0<k r\lesssim 1$ the approximation holds
\begin{equation}
K_{0}(kr)\sim -\ln{kr}.
\end{equation}
To capture the impact of the peaks we crudely approximate $\psi^2(k;0)$ with a zero-valued function everywhere except for the segment $(k_{0},k_{0}+\varepsilon)$ where the function has  amplitude $\tilde{\psi}$. Hence,
\begin{equation}\label{log}
\int_{0}^{\infty} dk K_{0}(kr)\psi^{2}(k;0)\approx\tilde{\psi}^{2}\int_{k_{0}}^{k_{0}+\varepsilon} dk \ln kr .
\end{equation}
 Up to an additive constant this integration yields
\begin{equation}
\phi(r)\sim\tilde{\psi}^{2}\varepsilon\ln(k_{0}+\varepsilon)r,
\end{equation} 
with $k_{0}=0$ as an admissible value. It is the sought-after voltage drop charateristic to two-dimensional currents with $\tilde{\psi}^{2}\varepsilon$ setting the inverse of the sheet conductivity parameter
\begin{equation}
\sigma_{2}=\frac{1}{2\pi\tilde{\psi}^{2}\varepsilon \sigma(\infty)}.
\end{equation}
Such a structure of the  resulting conductivity profile is retained if more realistic peak models (Lorentzian, Gaussian) are considered. Note, that peaks located even at large $k_{0}$ can induce logarithmic voltage bias at small, experimentally unattainable distances according to the scaling induced by $r$ in eq.~(\ref{integ}). \\

The function $\psi^{2}(k;0)$ is  determined by  $\sigma(z)$.  It is a challenging task to characterize links between the peak characteristics and the conductivity profile. According to our numerical observations, $\psi^{2}(0;0)$ is a minimum if the bulk value of conductivity is approached from below and maximum in the opposite case. This seems to be due to the (zero) resonance/anti-bound state character of  $\sqrt{\sigma(z)}$ in  eq.~(\ref{1dim}). Further, the potentials $V(z)$ immediately suggest existence of shape resonances in the theory. Indeed, peaks correspond to solutions that look like localized eigenfunctions in the  potential well.   Note, that the usual language using transmitted and reflected waves is not suitable  in the case of real-valued field. However, this does not appear to be a tough difficulty, see e.g.~\cite{Barry}.  \\

To conclude, we have demonstrated how the two-dimensional surface currents emerge within the classical current equation. They can be seen at finite distances only. Surprisingly, both the two-dimensional and three-dimensional currents at sufficiently large distances are governed by very few parameters: position, amplitude, and width of the pertinent peak of $\psi^{2}(k;0)$ function. Their knowledge allows calculation of the sheet-conductivity and the effective surface conductivity.

\acknowledgements
Founding for this research has been provided by EC under the Large-scale Integrating Project in FET Proactive of the 7th FP entitled ``Planar Atomic and Molecular Scale devices" (PAMS). 
\appendix
\section{Boundary conditions}\label{app}
In the paper~\cite{PRB} it was noted that the condition $\tfrac{d}{dz}\sigma(0)\neq 0$ may result in the large $r$ asymptotic behavior $\phi(r)  \sim r^{-3}$. This could be the case if $\psi(0;0)=0$, as suggested by the (generalized) eigenfunctions of   eq.~(\ref{1dim}) with zero potential. These solutions read $\{cos(kz +\varphi(k))\}_{k>0}$, where 
\begin{equation}
\varphi=-sign\left(\frac{d\sigma(0)}{dz}\right)\arccos{\frac{k}{\sqrt{k^{2} + \left(\frac{1}{2\sigma(0)}\frac{d\sigma(0)}{dz}\right)^{2}}}}.
\end{equation}
It is evident that these functions  vanish for $z=0$ and $k\to 0$. This is not the case if any relevant potential is taken into account. Note that  function $\sqrt{\sigma(z)\sigma^{-1}(\infty)}$ satisfies the boundary condition~(\ref{boundaryTransformed}) and solves eq.~(\ref{1dim}) with $ k=0$;  factor $\sigma^{-1}(\infty)$ ensures the correct normalization. As such, it locally describes how the solutions for small $k$ behave near $z=0$. Hence, $\psi^{2}(0;0)= \sigma(0)\sigma^{-1}(\infty)$ and the usual asymptotic behavior $\phi(r)\sim(\sigma(\infty)r)^{-1}$  holds for $r\to \infty$. The arguments given in~\cite{PRB} for $k\to\infty$ remain  valid in the case at hand as $\cos^{2}\varphi(k)\to 1$ for $k\to\infty$.\\


\begin{thebibliography}{9}
\bibitem{popular} S. Hasegawa, F. Grey, Surf. Sci. 500, 84 (2002); Ph. Hofmann and J. W. Wells, J. Phys.: Condens. Matter 21, 013003 (2009); An-Ping Li, K. W. Clark, X.-G. Zhang and A. P. Baddorf, Adv. Funct. Mater. 23,  2509 (2013).
\bibitem{APL} M. Wojtaszek, J. Lis, R. Zuzak, B. Such and M. Szymonski, Appl. Phys. Lett. 105, 042111 (2014)
\bibitem{JAP}  M. Wojtaszek, R. Zuzak,  S. Godlewski, M. Kolmer,  J. Lis, B. Such, and M. Szymonski, J. Appl. Phys. 118, 185703 (2015)
\bibitem{2d1} C. M. Polley, W. R. Clarke, J. A. Miwa, M. Y. Simmons, and J. W. Wells, Appl. Phys. Lett. 101, 262105 (2012)
\bibitem{2d2} C. Liu, I. Matsuda, S. Yoshimoto, T. Kanagawa, and S. Hasegawa, Phys. Rev. B 78, 035326 (2008).
\bibitem{2d3} L. Barreto, L. K{\"u}hnemund, F. Edler, Ch. Tegenkamp, J. Mi, M. Bremholm, B. B. Iversen, Ch.~Frydendahl, M. Bianchi, and Ph.~Hofmann, Nano Lett. 14, 3755 (2014)
\bibitem{PRB} J. Lis, M. Wojtaszek, R. Zuzak, B. Such, and M. Szymonski, Phys. Rev. B 92, 035309 (2015)
\bibitem{Calderon} G. Uhlmann, Inverse Problems 25, 123011 (2009) 
\bibitem{EIT} L. Borcea, Inverse Problems 18, R99 (2002)
\bibitem{Landau} L. D. Landau and E. M. Lifshitz, \emph{Electrodynamics of Continuous Media}  Pergamon Press Ltd., 1963, Chap. III
\bibitem{Rogala} M. Rogala, Z. Klusek, C. Rodenb{\"u}cher, R. Waser, and K. Szot, Appl. Phys. Lett. 102, 131604 (2013)
\bibitem{Monch} W. M{\"o}nch, \emph{Semiconductor Surfaces and Interfaces}  Springer, Berlin, New York, 2001.
\bibitem{Barry} B. Simon, J. Funct. Anal. 178, 396 (2000)

\end{thebibliography}
\end{document}